\documentclass[preprint]{aastex63}%
\usepackage[whole]{bxcjkjatype} 
\usepackage{afterpage}
\usepackage{graphicx}
\usepackage[normalem]{ulem}
\usepackage{lineno}
\usepackage{here}
\usepackage{xcolor}
\usepackage{multirow}
\usepackage{array}
\usepackage{booktabs}
\usepackage[utf8]{inputenc}
\newcommand{\jm}[1]{{}}
\newcommand{\jmold}[1]{{}}

\newcommand{\UVC}{\textit{u-v}~}

\newcommand{\virgoa}{M\,87\hspace{0.7mm}}

\newcommand{\DEG}{^{\circ}}

\newcommand{\etal}{et al.}

\newcommand{\RS}{$R_{\rm S}$}

\DeclareRobustCommand{\erase}{\bgroup\markoverwith{\textcolor{red}{\rule[.5ex]{2pt}{0.4pt}}}\ULon}
\def\jm#1{{\bf[#1 -- JM]}}

\received{15, September, 2023}
\accepted{27, January, 2024}
\submitjournal{ApJL}
\shorttitle{Jet and resolved features of M\,87 observed with EHT~in 2017}
\shortauthors{Miyoshi et al.}
\begin{document}
\title{The jet and resolved features of the central supermassive black hole \\ of M~87 observed with EHT~in 2017\\-- Comparison with the GMVA 86~GHz results}
\correspondingauthor{Makoto Miyoshi}
\email{makoto.miyoshi@nao.ac.jp}
\author[0000-0002-6272-507X]{Makoto Miyoshi}
\affil{National Astronomical Observatory, Japan, 2-21-1, Osawa, Mitaka, Tokyo, Japan, 181-8588}
%
%
\author[0000-0003-2349-9003]{Yoshiaki Kato}
\affil{Japan Meteorological Agency: 3-6-9 Toranomon, Minato City, Tokyo 105-8431, Japan, e-mail: yoshi\_kato@met.kishou.go.jp}
%
\author[0000-0002-0411-4297]{Junichiro Makino}
\affil{Department of Planetology, Kobe University, 1-1 Rokkodaicho, Nada-ku, Kobe, Hyogo 650-0013, Japan, e-mail: makino@mail.jmlab.jp}
\author[0000-0001-8185-8954]{Masato Tsuboi}
\affil{School of Science and Engineering, Meisei University, 2-1-1 Hodokubocho, Hino City, Tokyo 191-0042, Japan, e-mail: masato.tsuboi@meisei-u.ac.jp}
\nocollaboration{4}
\begin{abstract}
\virgoa is the best target for studying black hole accretion and jet formation. Reanalysis of the EHT public data at 230~GHz shows a core-knots structure at the center and jet features~\citep{Miyoshi2022a}.
We here compare this with the new results of GMVA at 86~GHz showing a spatially resolved central core~\citep{Lu:2023}.
There are similarities and differences between the two. 
At 86~GHz, "two bright regions" are seen on the ring in the core.
"Core-Knot-Westknot", triple structure in the 230~GHz image shows apparent appearance of two peaks similar to the "two bright regions" when convolved with the GMVA beam.
This similarity suggests that both frequencies reveal the same objects in the core area.
Protrusions are observed on both the south and north sides of the core at both frequencies, becoming prominent and wing-like at 230~GHz.
The 86~GHz image shows a triple ridge jet structure, while the 230~GHz image shows only a bright central ridge with two roots.
Both frequencies show a shade between the core and the central ridge.
To detect the faint features from the EHT2017 data, we found that the use of all baseline data is essential. Using all including the ultrashort baseline data, revealed the jet and faint structures. Without the ultrashort baselines, these structures were not detectable. The lack of detection of any faint structures other than the ring in the \virgoa data by the EHTC is presumably due to the exclusion of ultrashort baselines from their analysis.
\end{abstract}
\section{Introduction}\label{Sec:Intro}
The supermassive black hole in \virgoa~is the best to study the origin of the jet because it has the largest apparent angular size for black holes with strong jets, due to the relatively small distance (16.7~Mpc;~\cite{Mei07}) and large mass ($6.1 \pm~0.4 \times 10^9M_{\odot}$;\cite{G2011} 
\footnote{The mass of~black hole~in~\virgoa ~is still controversial. A mass of $M_\mathrm{BH} = (3.5^{+0.9}_{-0.7}) \times 10^9\ M_\odot$ (68~$\%$ confidence) is obtained from gas dynamics~\citep{WBHS2013}.}), implying $1~R_{\rm S}$ = 7$~\mu\rm as$.
 
The 230~GHz Event Horizon Telescope (EHT) observations by the EHT Collaboration (EHTC) revealed a ring image consistent with the predicted shadow size of \virgoa~black hole~\citep{RefEHT1-6}.
The EHTC papers provide detailed descriptions of the analysis. However, the point spread function (PSF) of the EHT array is missing, especially the position and intensity of the sidelobes relative to the main beam, which is crucial for identifying potential imaging artifacts. 
This led \cite{Miyoshi2022a} (hereafter MKM22) to reanalyze the EHT public data, finding that the reported ring image closely resembled the PSF structure. MKM22 suggested that the EHTC ring image was likely an artifact resulting from the EHT's sparse \UVC coverage and sampling bias, particularly the absence of spatial Fourier components corresponding to the diameter of the EHTC ring image, coupled with imaging analysis using very narrow field-of-view settings. All other independent ring results are also suggested to have the same origin~\citep{Miyoshi2022b}.
Additionally, MKM22 reported an image more consistent with the data than the EHTC ring image. It shows a central structure, appearing as a core with two knots (hereafter C-K-W structure; Core-Knot-Westknot), and a large brightness distribution area aligning with the previously known jet.
However, there were no previous or independent observations of this C-K-W structure, so its validity was determined solely by its better agreement with the data compared to the EHTC ring images.

Recently, \cite{Lu:2023} (hereafter Lu23)~ reported the GMVA image of \virgoa, showing that the core is also spatially resolved at 86~GHz. 
~This is due to the improved spatial resolution resulting from the addition of the phased Atacama Large Millimeter/submillimeter Array (ALMA) and the Greenland Telescope (GLT) to the GMVA network. Default beam
\footnote{
In this paper, `Default beam' refers to the shape obtained by an elliptical Gaussian fit to the main beam shape in the PSF structure.}
with a full width at half maximum (FWHM) of ~$79\times 37~\mu \rm as$ and a PA of $-63^{\circ}$ is obtained. 
In the core, Lu23 found "two bright regions" of emission oriented in the north-south direction at the base of the northern and southern ridges of jet. 
Lu23 interpreted this core configuration as a ring structure 
 with a diameter of $64^{+4}_{-8}~\mu\rm as$~ 
($8.4^{+0.5}_{-1.1}$~\RS\footnote{Their assumptions include a black hole mass of $M = 6.5 \times 10^{9} M_{\odot}$ and a distance of 16.8~Mpc.}), which is about $50\% $ larger than the diameter measured by the EHTC. 
Lu23 also show that a triple-ridge jet structure emerges from the spatially resolved core extending westward~( Lu23's~ Figure 1 (a)).
This structure features sharp gaps of emission between the ridges, connecting to larger-scale structures ($\geq 100$~\RS) previously observed~\citep{Kim2018,Walker2018}.

 Indeed,~Lu23~ presents high-resolution images that are comparable to those obtained by MKM22.
In this paper we compare the images from MKM22 with those in Lu23~ .
Section~\ref{Sec:ourreduction} 
details the additional processing applied to the images used in this study.
Since the data calibration and imaging process have been thoroughly described in MKM22, 
we do not repeat them here.
Our imaging results are presented in Section~\ref{Sec:comparison}. Section~\ref{Sec:disc} includes our discussion of these results, and a summary is provided in Section~\ref{Sec:CR}. 
The discussion of the effect of the ultrashort baselines on the images is presented in Section~\ref{Sec:disc}, and the simulation results are provided as evidence in Appendix~\ref{Sec:simulation}.
\vspace{1cm}

\section{The 230~GHz images for comparison}\label{Sec:ourreduction}
The 230~GHz image presented in this paper is based on the results of the MKM22 analysis. Refer to MKM22 for details on data calibration, the method used to obtain the final image, and the reliability
\footnote{
The EHT public data differ from standard archived VLBI data, such as those from the VLBA, primarily due to compression in both time and frequency directions. This compression, particularly the integration of frequency channels into a single channel, precludes critical procedures like bandpass corrections and fringe searches in the delay and delay-rate directions. Consequently, self-calibration is the only feasible calibration method.}.
However, for the wide area image , which includes the jet, we continued our analysis and were able to obtain
the jet structure near the core.
 These results are then utilized for comparison about the large-scale structure.

\subsection{Core image of M~87 for comparison}
\label{Sec:230Gcore}
MKM22 found the C-K-W structure, triple compact emissions at the core, about $100~\mu \rm as$ in size and encircled by fainter emissions. 
MKM22 show two images, one from the first two days' data, and the other from the last two days' data. For the comparison, we created an averaged image of them.

\subsection{Large-scale image for comparison}
\label{Sec:230jet}
In MKM22, phase-only self-calibrations were used to find the final images. This time, to more clearly show the initial jet structure near the core, we created a new image. 
For data calibration, we used the amplitude and phase self-calibration solutions shown in Figures 5 and 6 in MKM22. After applying these solutions as calibration to the data, we made 29 CLEAN maps with different CLEAN loop gains~($1\times 10^{-3}$~to $0.1$). Other parameters for the CLEANs are the same as those shown in Table 4 in MKM22. We selected the CLEAN result with the smallest residuals in closure phase and amplitude as the "best" image.
 
 The EHT2017's \UVC coverage could potentially lead the CLEAN algorithm to introduce artifact structures as CLEAN components~(MKM22).
 
To minimize these artifacts, we created restricted images using selected CLEAN components from the "best" image. We then assessed the consistency between these restricted images and the data by analyzing the residuals of the closure quantities (Table~\ref{tab:residuals}).
 
~Judging from the residuals, the 'C only' 
and the 'C-K-W area' alone are insufficient to explain the observed data; the 'Concentrated region' 
and the components that appear to be part of the jet extending up to 8~mas to the west must also be included to fully explain the data. Then, we selected the CLEAN components of case (2) in Table~\ref{tab:residuals} as representing the 'final' image in Figure~2\\

\begin{table}
\centering
\begin{tabular}{lrr}
\toprule

 Case ~~~~~~~~~~~~~~&Residuals of Closure Amplitude~&~Residulals of Closure Phase\\
~~~~~~~~~~~~~~~~~~~&~~~~~~~~~(Normalized) ~~~~~~~~~&~~~~~~~~~~($^{\circ}$)~~~~~~~~~~~~\\ 
\midrule
(1) All CLEAN components     &   $~0.496\pm~~2.114$&$~1.2\pm52.2$\\
(2) Within $95\%$ of each BOX's radius&$~0.410\pm~~1.908$ & $~0.3\pm52.8$\\
(3) Concentrated region& $~0.479\pm~~2.484$ &$-0.4\pm53.4$\\
(4) C-K-W area&$~2.498\pm~14.783$&$-0.3\pm61.2$\\
(5) C only&$13.802\pm103.374$&$20.9\pm74.4$\\
\bottomrule
\end{tabular}
\caption{
Closure residuals.
 Case (1) is all components obtained by CLEAN,
 Case (2) is the components within $95~\%$ of the radii of the circles of the eight "BOX" used in CLEAN,
 Case (3) is the components belonging 
to the 0.7~mas area around the origin, where there is a C-K-W structure and a less bright CLEAN component surrounding it.
 Case (4) is the components within a narrow range including the three components of "Core", "Knot", and "West" found in MKM22, and
 Case (5) includes only the "Core" components
 almost a point source in the center. 
Case (2) shows the smallest residuals considering both closure amplitude and closure phase. 
All calculations are based on 180~seconds of integrated data.
}\label{tab:residuals}
\end{table}

One important imaging factor must be noted: We observed a critical aspect regarding the detection of jet features.
Using all available baseline data revealed the jet features. The data include the SMA-JCMT baseline ($154$ to $159~\rm m$) and the ALMA-APEX baseline ($2028$ to $2503~\rm m$), both in projected baseline length.
 Such ultrashort baseline data have not been obtained in other VLBI networks. We attempted to run CLEAN using only long baseline data, excluding such ultrashort baseline data. The results showed that neither jet features nor weak components near the center could be detected.
The ultrashort baselines are significant contributors to the detection of faint structures from the EHT2017 public data\footnote{
AIPS IMAGR issue: When CLEAN maps are generated by IMAGR from EHT2017 data, including the ultrashort baselines, the resulting CLEAN maps often show flux densities that are orders of magnitude higher than the observed visibility data.
It is obvious that such maps are not consistent with the observed data.
On the other hand, the CLEAN components alone always show the appropriate flux density.
We conclude that something is wrong in the process of combining the last residual image with the CLEAN components to produce the CLEAN map.
The task IMAGR in AIPS may not be able to properly handle data that contains both extremely short and very long baselines.
Therefore, for analysis in MKM22 and this paper, we use the set of CLEAN components instead of the CLEAN map.
}.
 Appendix~\ref{Sec:simulation} describes simulations that support this.
\vspace{10mm}
\section{Comparison of the Imaging results\label{Sec:comparison}}
In this section, we present our images of \virgoa at 230~GHz and compare them with those at 86~GHz from Lu23.

Differences in spatial resolution can make image comparisons difficult and should be adjusted and compared.
 Fortunately, the MKM22 image has higher spatial resolution, the resolution can be reduced.
For adjustment, we convolved Lu23's restoring beams to the MKM22 image to create images of the same spatial resolution as Lu23's. However, because the resolution of Lu23's super-resolution image is not noted, we assumed it to be a $20~\mu \rm as$ circular Gaussian and compared the image.
 
All 86~GHz images used here are taken directly from Lu23 . They are used as they appear in the original paper, except that some have been enlarged and trimmed for comparison, and some of the contour colors have been altered.

In VLBI imaging, the accuracy of absolute coordinates for imaged objects is not as precise as their spatial resolution.
Here, the alignment of the two images was determined by adjusting the C-K-W area to fall within Lu23's "two bright regions" in the highest resolution image. The alignment of other resolution overlays is also defined by this.
\clearpage
\subsection{Core Region}\label{Sec:core}
\begin{figure}[H]
\begin{center} 
\epsscale{1.165}
\plotone{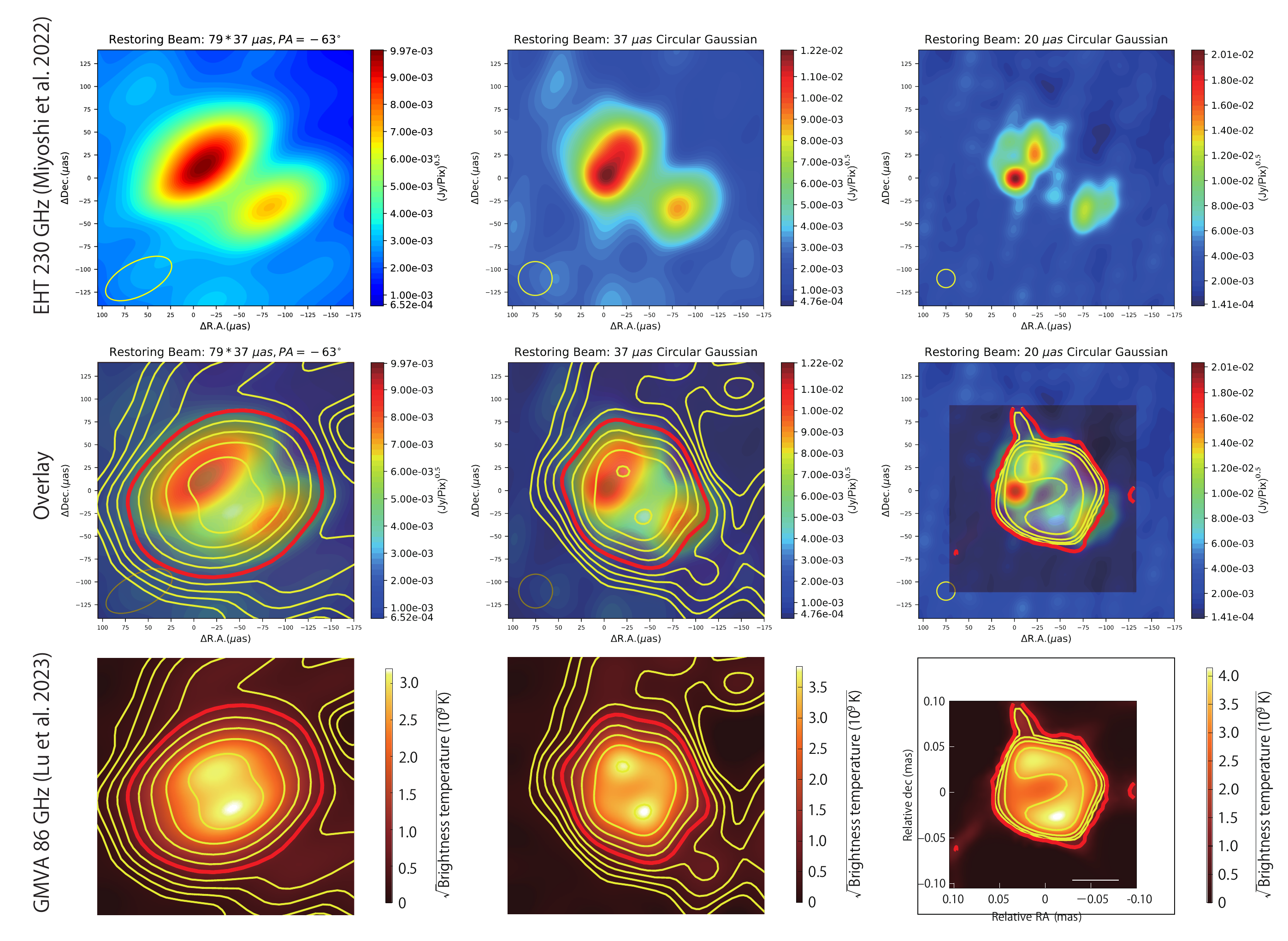}
\end{center} 
\vspace{-4mm}
\caption{
Comparison of core structures between the GMVA2018 86~GHz results (Lu23) and the EHT2017 230~GHz data reanalysis~(MKM22).
Top: The average images of the C-K-W area obtained from the EHT 230~GHz data reanalysis (see MKM22 for details). 
From left to right, images convolved by the restoring beams of the GMVA default beam, $37~\mu \rm as$ circular Gaussian beam, and $20~\mu \rm as$ circular Gaussian beam, respectively.
The brightness unit is Jy/pixel, where $1~pixel = 1.5~\mu \rm as$.
Middle: Overlays of the EHT 230~GHz images (top panels) and those corresponding to the same spatial resolutions from Lu23's GMVA 86~GHz observations, shown in the bottom panels.
The levels of Lu23's contour lines highlighted in red correspond from left to right to approximately $4.4~\%$, $5.3~\%$, and $4~\%$ of their respective peak brightness.
Bottom: Lu23's GMVA 86~GHz results,~modified from their Figure 1 (a), (b), (c).
 Lu23's contour lines are drawn at intervals representing a doubling in brightness. 
See~ Lu23 for details.
}\label{Fig:coreimages}
\end{figure}

Figure~\ref{Fig:coreimages} shows the comparisons of the core images.
At $20~\mu \rm as$, the resolution for the EHT 230~GHz observations, features C, K, and W appear completely separated~(top right panel). 
 However, with lower spatial resolution, C and K appear to be connected as one bright region.
Thus, at these lower resolutions, they appear as two distinct peaks (remaining two top panels).
At all resolutions, the sizes of the cores with "two bright regions" at 86~GHz and the C-K-W area at 230~GHz are identical.

In the lowest resolution images~(left side panels), the C-K-W area falls within the $5^{th}$ contour line in the image of Lu23 ($8~mJy/beam~level$), namely within the area of the $4.4\%$ level of Lu23's peak intensity~($180~mJy/beam$).
For clarity, the corresponding contour lines are shown in red in the middle and bottom panels of Figure~\ref{Fig:coreimages}. 
The same approach is applied below. 

In the middle resolution images (central panels), the C-K-W area also falls within the $5^{th}$ contour line in the image of Lu23 ($6.4~mJy/beam~level$), namely also within the area of the $5.3\%$ level of the peak intensity~($120~mJy/beam$).

Because the noise level of Lu23's images is uncertain, the image reliability cannot be determined by the signal-to-noise ratio. 
Nevertheless, the brightness at the lowest contour level would likely exceed the minimum detection level. If that is the case, then the brightness in the area above the $5^{th}$ contour line is 16 times that of the lowest one, suggesting the area is certainly within the core. This area includes both the "two bright regions" and the C-K-W area. The C-K-W area is very likely to be the same as the "two bright regions".

As for the 86~GHz highest resolution image, assuming a resolution of $20~\mu \rm as$, we find that the C-K-W area and Lu23's core are almost the same size. 
(In that panel of Figure~\ref{Fig:coreimages}, the contour line at the $4~\%$ level of the peak are highlighted in red.)

~The separations between the "two bright regions" at 86~GHz are all larger than those of the two peaks at 230~GHz.
~However, this difference can be interpreted as due to the time variability of the core structure, since there is about one year's gap between the two observations.
In fact, the C-K-W structure at 230~GHz shows a position shift within a week; the proper motions of K and W correspond to $\sim 0.3~\rm mas/yr$ ($\sim0.1~\rm c$) (MKM22).

If the motions are real, the one-year gap in observations naturally results in an image of GMVA2018 different from that of EHT2017.
 
The internal structure of the core is indeed resolved as at least two discrete regions within it, but further observations are needed to determine their true nature.

Note that the flux density in the C-K-W region within $240~\mu \rm as$ diameter is 446~mJy, not much different from that of the Lu23'core (562-591~mJy)~\citep{Lu:2023s}.

\subsection{Large-scale\label{Sec:large}}
 Here we compare our~ "final" image with~ Lu23's large-scale image in their Figure 1 (a). 
 In Figure~2, we show our image and the overlay with that of Lu23.
The observational epochs of EHT2017 and Lu23's GMVA are about 1 year apart, so a detailed comparison is not meaningful due to the existence of time variation. However, here, a comparison is attempted. 
 We describe the 230~GHz image characteristics, with corresponding features of Lu23's image noted in parentheses.

The overall brightness distribution extends from 2~mas east to 8~mas west~(Figure~2~(a)). 
However, the brightest region is from the core to about 0.5~mas west. [A similar range distribution when limited to its bright areas more than $1~\%$ level of the peak brightness shown by orange contours in 
Figure~2~(c).]

The brightest area is the core. 
 [The core is also the brightest in Lu23's observations].
The core appears to have two peaks due to the low spatial resolution of Lu23's default beam, as shown in Figure~\ref{Fig:coreimages} (b). [The "two bright regions" are seen.]
The outer edge structure of the core has wing-like protrusions on the south and north sides of the core.
[Lu23 also shows protrusions on the north and south sides of the core.]
Of these, the south wing is brighter than the north one and has a spread of about twice as large, or about 0.2~mas.
The south wing is considered realistic based on the results of the closure residual study -- the closure residual becomes larger when the south wing is removed.
The angle formed by the north wing, core, and south one is about $120\DEG$. 
[The angle for the 86~GHz case is also larger than $90\DEG$.]

 Next, the jet structure is discussed.
 The bright central ridge A  extends from between the two bright peaks of the core to about 0.5 mas to the west. 
[The central ridge at 86~GHz extends about 1 mas.] 
The central ridge B extends westward from the south wing. 
Two central ridges,~ originating from independent roots, appear to merge into a single central ridge.
(It is unlikely that central ridge B corresponds to Lu23's south ridge. If so, it should move away from the central ridge.)

There is a shade, a relatively dark area bounded by the core and the two central ridges. Such shades are also seen in the GMVA2014 observations~\citep{Kim2018} but not seen in GRMHD simulations~\citep{Cruz-Osorio:22}.
 [A shade, albeit narrower than that at 230~GHz, is also present between the core and the central ridge.]
 
The central ridge is bright while the presence of ridges on either side is uncertain. 
So-called edge-brightened jet structure is absent. 
[There are distinct and independent triple ridges on each side and in the center. The ridges on either side are brighter than the central ridge. However, at 0.25~mas from the core, there appears to be no difference in brightness between the three].
Note that the ridges are not continuous at either frequency, but are formed by lines of separate knots.

An eastward-extending structure, maybe a counterjet, is observed. The presence of counterjets have been suggested from previous 43, and 86~GHz observations.
[Lu23 found no sign of counterjets.]
 
\noindent
\begin{minipage}{\linewidth}
    \centering
    \includegraphics[width=\linewidth]{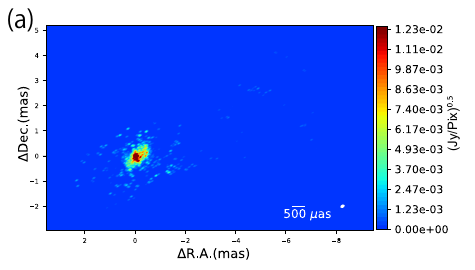} 
\end{minipage}
\noindent
\begin{minipage}{\linewidth}
    \centering
    \includegraphics[width=\linewidth]{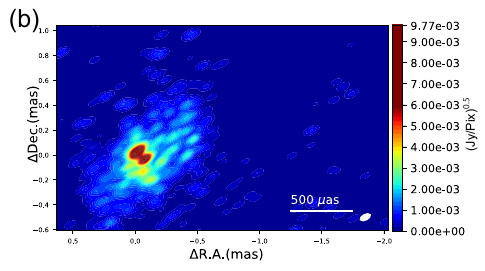} 
\end{minipage}
\clearpage 
\noindent
\begin{minipage}{\linewidth}
    \centering
    \includegraphics[width=\linewidth]{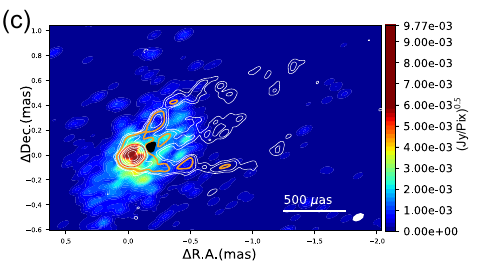} 
\end{minipage}
\noindent
\begin{minipage}{\linewidth}
    \centering
    \includegraphics[width=\linewidth]{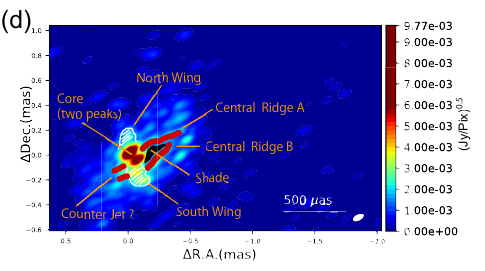} 
\end{minipage}
\begin{minipage}{\linewidth}
{\bf Figure 2:} ~Our image of \virgoa obtained from the EHT2017 230~GHz observations (the last two days of data). The image was constructed from the CLEAN components of Table~\ref{tab:residuals}, Case (2).~ Lu23's default beam, an elliptical Gaussian with an FWHM of $79\times 37~\mu \rm as$ and a PA of $-63\DEG$, is used as the restoring beam, which is shown as a white ellipse in each image.
(a) The full view of \virgoa. The brightness unit is Jy/pixel, the pixel size is $6~\mu \rm as$.
(b)-(d) The enlarged images of the core region. The brightness unit is Jy/pixel, the pixel size is $1.5~\mu \rm as$.
(c) Overlays with the contours of Lu23's image shown in their Figure 1 (a). The minimum contour level is $0.5~mJy/beam$, about $0.28~\%$ of the peak brightness of Lu23's image. Subsequent contour levels represent a doubling of brightness. The orange contours correspond to about $1.1~\%$ of the peak, and the red contour corresponds to about $4.4~\%$ of the peak. 
The blackened region indicates "shade", which is relatively darker than the surrounding regions between the core and the central ridge.
(d) Naming of each region in our image is shown.
\label{Fig:largeimages}
\end{minipage}


\section{Discussion\label{Sec:disc}}
 We begin our discussion with the core structure of \virgoa, comparing the images from our analysis with those of~Lu23. 
We then explore the imaging performance of using the EHT's ultrashort baseline data.
 
\subsection{The structure of the core region in M 87}\label{Sec:discA}
 
\begin{itemize}
\item The inner structure of \virgoa's core was first revealed by EHT and GMVA observations. The C-K-W structure~(MKM22) and "two bright regions"~(Lu23) were detected, respectively, and since both are the same size, we can consider that they are the same thing. 
MKM22 interpreted them as a true core and knots. Lu23 interpreted them as a bright part of the circum-nuclear ring.
 
The hour angles of the line connecting the two bright peaks in the reduced-resolution 230~GHz images differ from those observed at 86~GHz. The difference might suggest the orbital motion of these bright objects. Jet precession in \virgoa has been observed through VLBI~\citep{Walker2018,Cui:23}. Considering the observed positional changes of these two objects in the core, it may be hypothesized the existence of a supermassive black hole binary. The orbital motions in such system could potentially be a contributing factor to the jet's precession.
 
\item 
Regarding the outer shape of the core, both observations show protrusions on the south and north sides.
Especially at 230~GHz they are prominent and appear to be wing-shaped.
The size of the protrusions is about $80~\mu \rm as~(12~R_{\rm S})$ at 86~GHz and $170~\mu \rm as~(25~R_{\rm S})$ at 230~GHz.
It is not clear whether they are outflows from the center that are distinct from the central jet, e.g., the glowing wind from the accretion disk, or, on the contrary, inflows to the center. 
\item Lu23 found a triple ridge jet structure extending from the core, which are completely independent. 
The central ridge-like structure was suggested at 43~GHz, but is~ now~ clearly visible at 86~GHz. 
However, the south and north ridges are brighter than the central ridge.
At 230~GHz, which is three times higher frequency, the existence of the south and north ridges becomes ambiguous and the central ridge is dominant. 
In addition, the central ridge at 230~GHz has two ridges at its root that appear to merge into a single ridge. 
This picture recalls the helical flow of jets near the core suggested by~\cite{Walker2018}, based on 17 years of 43~GHz VLBA observations. While the south and north ridges tend to be brighter at lower frequencies, the central ridge is relatively brighter at higher frequencies. 
 The central ridge may have a different physical origin for its emission than the south and north ridges.
\item
At both frequencies, the core and central ridge are not continuous, and there is a shaded area between the core and central ridge, which is also seen in the GMVA2014 results~\citep{Kim2018}. The shade is more extensive at 230~GHz. 
The length of the shade reaches $240~\mu \rm as~(34~R_{\rm S})$ at 230~GHz, and 
$60~\mu \rm as~(8~R_{\rm S})$ at 86~GHz.
This shade may be closely related to the emission mechanism of the jet, especially that of the central ridge.
\end{itemize}
~Additionally, Radio Astron's 22~GHz observations~\citep{Kim2023} show a 0.36~mas core extending north-south, with a bright south side. This size is similar to that of the core region including both wings at 230~GHz; the south wing is brighter and larger at 230~GHz, consistent with the 22~GHz core's~brightness distribution. 
Also a shade between the core and the first knot are seen.
 \subsection{The imaging performance using the EHT ultrashort baseline data.}\label{Sec:discB}
Compared to GMVA's 86~GHz image, EHT's 230~GHz images also detect an unexpectedly wide variety of features though the EHT2017 observations were made with only seven stations at five locations on the globe. It is reasonable to expect that they would not detect as many features as the GMVA2018 observations with more than ten stations. However, investigating the EHT2017 data, we found that the use of the ultrashort baselines contained in the data allows the detection of much fainter features than the bright core structure~(Sections~\ref{Sec:230jet} and Appendix~\ref{Sec:simulation}). 

Most interferometers are poor at detecting extended source structure, but the addition of single-dish data corresponding to zero-spacing information solves this problem~\citep{Wilner:94,Kurono:09,Plunkett:23}.
 
The role of the ultrashort baseline found here differs slightly from that above. The inclusion of near-zero spacing data improves the detectability of faint but very compact structures.
 
The EHT array will provide exceptional imaging capabilities, not only because of its highest spatial resolution, but also because of its ultrashort baselines, which other interferometers do not have.
 
The EHTC analysis was likely performed without the ultrashort baseline data
\footnote{
For example, the EHTC imaging pipeline using the DIFMAP
(\url{https://github.com/eventhorizontelescope/2019-D01-02};
\url{difmap/EHT_Difmap})~produces images after omitting the ultrashort baselines.
}. 
If so, this is one reason why the EHTC analysis could not detect the jet structure.

\section{Summary\label{Sec:CR}}
We performed comparisons between the \virgoa image from the recent 86~GHz GMVA observation~\citep{Lu:2023} and the 230~GHz image derived by applying the final self-calibration solutions shown in \cite{Miyoshi2022a} to the EHT2017 public data.
The resolved core at the 86~GHz is just in the size of C-K-W area found at 230~GHz, and both show two of bright spots when the spatial resolutions of the 230~GHz image are adjusted to those of 86~GHz images.
 Presumably both observations detected the same objects in the core.
 However, there is a difference in the spacings and hour angles between the two bright spots, presumably due to time variation.
 
The protrusions on the south and north sides of the core were observed at both frequencies. Whether these are inflows toward the black hole or outflows other than jets, such as disk winds from the accretion disk, is not yet known.
~At 86~GHz, both sides of the ridge structure observed at lower frequencies were visible, and the central ridge long suggested at lower frequencies was clearly seen.
At 230~GHz, the ridge structure on both sides of the ridge becomes unclear, and only the central ridge is clearly visible. 
The central ridge appears to originate from two distinct roots that coalesce to form a single central ridge. This may be related to the helical flow suggested in the 43~GHz observations.
At both frequencies there is a shade between the core and the central ridge.
If the gap is universally present, it could be a factor in determining the mechanism of jet generation.
At higher frequencies, the ridges on either side become darker, while the central ridge becomes brighter. 
The GMVA and EHT observations initially unveiled the core's inner structure and its jet connection.

~We noticed that the detection of the jet features from the EHT2017 public data depends on the utilization of the ultrashort baseline data it contains. 
The EHT is characterized by very long baselines and high frequency observations, which provide the highest spatial resolution. However, the array's ultrashort baseline is very powerful in detecting faint structures because it samples very low frequency spatial Fourier components.
 We successfully detected \virgoa's faint structures utilizing EHT's ultrashort baseline data.
\acknowledgments
We would like to thank Takahiro Tsutsumi and Rei Shinnaga-Furuya for providing us with information on the related imaging technique.
This work is supported in part by the Grant-in-Aid from the Ministry of Education, Sports, Science and Technology (MEXT) of Japan, No.19K03939.
We thank the anonymous referee for helpful comments that greatly improved the paper and the discussion.
This study owes much to the public archives of the EHT and the recent GMVA observations of Lu23,~\citep{Lu:2023}. 
We sincerely thank them for their invaluable contributions.
The Global Millimetre VLBI Array (GMVA) consists of telescopes operated by the Max-Planck-Institut fur Radioastronomie (MPIfR), IRAM, Onsala, Metsahovi Radio Observatory, Yebes, the Korean VLBI Network, the Greenland Telescope, the Green Bank Observatory (GBT) and the Very Long Baseline Array (VLBA). 
The VLBA and the GBT are facilities of the National Science Foundation (NSF) operated under cooperative agreement by Associated Universities. 
The GMVA data were correlated at the VLBI correlator of MPIfR in Bonn, Germany. 
We thank the EHT Collaboration for releasing the network-calibrated \virgoa data.
The EHT2017 observations of \virgoa were performed with the following seven telescopes.
ALMA is a partnership of the European Southern Observatory (ESO; Europe, representing its member states), NSF, and National Institutes of Natural Sciences of Japan, together with National Research Council (Canada), Ministry of Science and Technology (MOST; Taiwan), Academia Sinica Institute of Astronomy and Astro- physics (ASIAA; Taiwan), and Korea Astronomy and Space Science Institute (KASI; Republic of Korea), in cooperation with the Republic of Chile. The Joint ALMA Observatory is operated by ESO, Associated Universities, Inc. (AUI)/NRAO, and the National Astronomical Observatory of Japan (NAOJ). 
The NRAO is a facility of the NSF operated under cooperative agreement by AUI. 
APEX is a collaboration between the Max-Planck-Institut f\"ur Radioastronomie (Germany), ESO, and the Onsala Space Observatory (Sweden). 
The SMA is a joint project between the SAO and ASIAA and is funded by the Smithsonian Institution and the Academia Sinica. 
The JCMT is operated by the East Asian Observatory on behalf of the NAOJ, ASIAA, and KASI, as well as the Ministry of Finance of China, Chinese Academy of Sciences, and the National Key R\&D Program (No. 2017YFA0402700) of China. Additional funding support for the JCMT is provided by the Science and Technologies Facility Council (UK) and participating uni- versities in the UK and Canada. 
The LMT project is a joint effort of the Instituto Nacional de Astr\.{o}fisica, \.{O}ptica, y Electr\.{o}nica (Mexico) and the University of Massachusetts at Amherst (USA). 
The IRAM 30-m telescope on Pico Veleta, Spain is operated by IRAM and supported by CNRS (Centre National de la Recherche Scientifique, France), MPG (Max-Planck-Gesellschaft,   Germany) and IGN (Instituto Geogr\.{a}fico Nacional, Spain).
The SMT is operated by the Arizona Radio Observatory, a part of the Steward Observatory of the University of Arizona, with financial support of operations from the State of Arizona and financial support for instrumentation development from the NSF. Partial SPT support is provided by the NSF Physics Frontier Center award (PHY-0114422) to the Kavli Institute of Cosmological Physics at the University of Chicago (USA), the Kavli Foundation, and the GBMF (GBMF-947). 
\facilities{GMVA, EHT2017}

\software{AIPS~\citep{Greisen2003}}

\clearpage
\appendix
\section{simulation for testing the effect of the ultrashort baseline data}\label{Sec:simulation}
~We performed simulations to investigate the effect of the ultrashort baselines contained in the EHT2017 data on the imaging results. The simulations confirmed that by including the ultrashort baseline data, faint structures, namely the jet features and the relatively faint structures in the core, can be detected. Without using the ultrashort baseline data for imaging, it becomes hard to detect the faint structures, and it also becomes difficult to capture the accurate structures of the bright core.
~The simulated data from a single point source yielded a result image showing only one point, both with and without the inclusion of the ultrashort baseline data.
~For the next simulation, we created a simulated visibility dataset (spatial Fourier components) 
corresponding to the model image shown in the top panels in Figures~\ref{Fig:simulation1} and \ref{Fig:simulation2}, and performed CLEAN on the data using the AIPS task IMAGR with the same parameters used by MKM22. The resulting maps of the CLEAN components are shown also in Figures~\ref{Fig:simulation1} and~\ref{Fig:simulation2}, using the default beam of the GMVA as the restoring beam.

~First, results for the wide area image are presented in Figure~\ref{Fig:simulation1}. In the model image (top panel), brightness is concentrated at the origin, and features are distributed in the direction of the jet and in the opposite direction. Faint features are scattered about from 2.5~mas east to 7~mas west of the origin.
~From the simulated data, when the ultrashort baseline data are flagged out and then CLEAN is performed on these flagged data, only the central bright core is imaged and no other structures are captured (see bottom panel of Figure~\ref{Fig:simulation1}).~On the other hand, if CLEAN is run on all the data including the ultrashort baselines, faint components present in the model image are detected.

~Second, results for the central core structure are shown in Figure~\ref{Fig:simulation2}.
CLEAN using data without the ultrashort baseline only detects the brightest of the two central peaks in the model image (see bottom panel of Figure~\ref{Fig:simulation2}).
~On the other hand, CLEAN using all data including the ultrashort baseline data accurately reproduces the core structure of the center (see middle panel of Figure~\ref{Fig:simulation2}).

The ultrashort baseline data appears to increase the dynamic range of the resulting image.
~As above, the results including the ultrashort baseline data largely reproduce the model image.
 However, when the ultrashort baseline data are excluded, the reproduction is not accurate.
If the EHTC excluded ultrashort baseline data from their analysis, faint jet structures could not be detected and the artifact ring image might appear at the core.

Note, however, that there are quite a few faint features that are not present in the model image. As estimated in MKM22, the resulting image (set of CLEAN components) contains brightness distributions that should not be there. (Compare the top and middle panels of Figure~\ref{Fig:simulation1}).

These are probably due to the presence of sidelobes in the PSF of the EHT2017 array that are comparable in strength to the main beam. 
Non-real brightness peaks appear as spurious peaks from the real brightness 
due to convolution with the PSF, and it is believed that these were picked up by the CLEAN algorithm.
\addtocounter{figure}{1}
\begin{figure}[H]
\begin{center} 
\epsscale{0.7}
\plotone{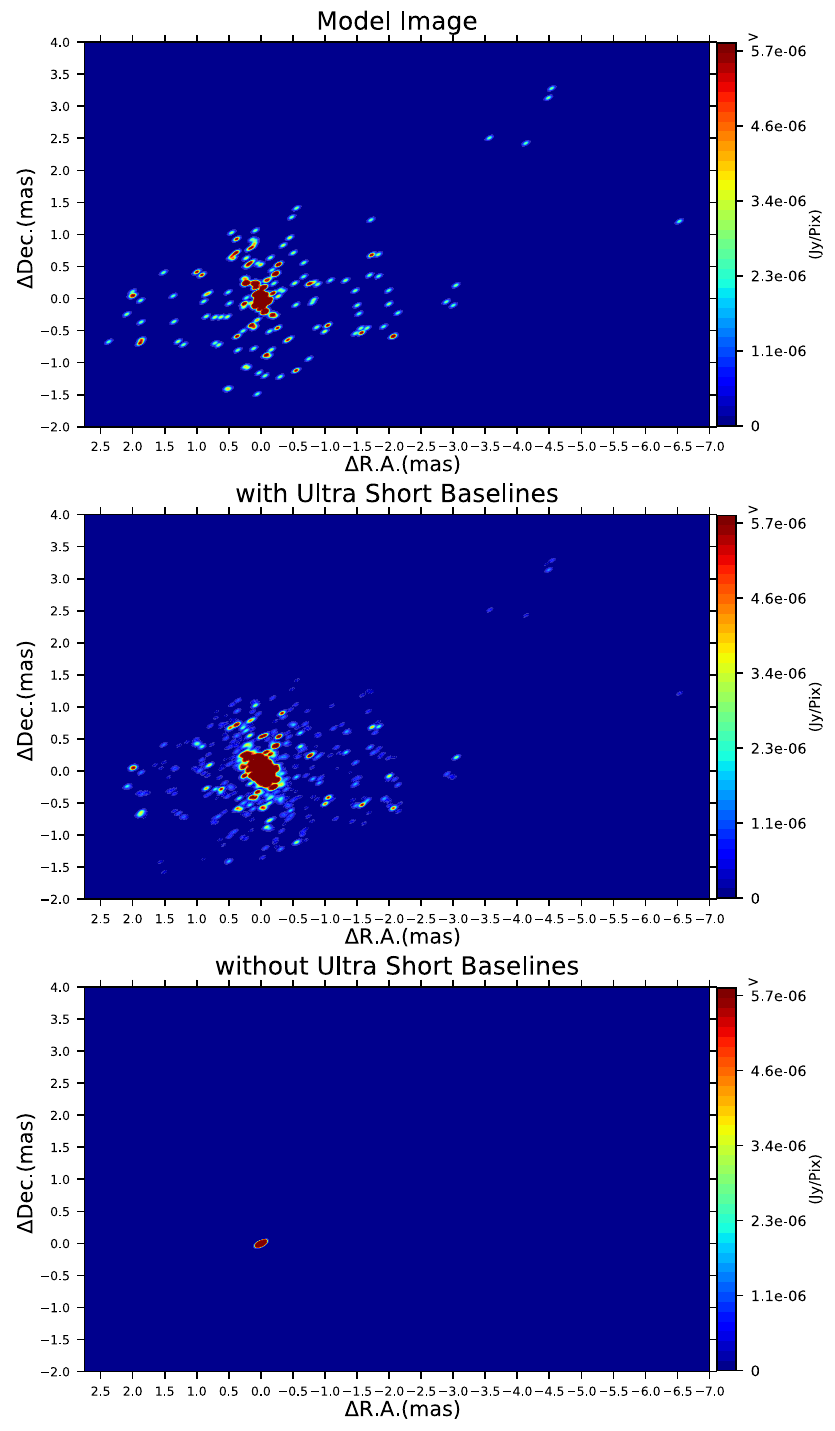}
\end{center} 
\vspace{-4mm}
\caption{
Simulation for the performance of the data from the ultrashort base lines (I).
Shown is a large area. The top panel shows the model image, and the middle panel shows the result when all data including the ultrashort baseline data are used.
The bottom panel shows the result when the ultrashort baseline data are omitted and only the remaining long baseline data are used. The brightness distributions are generated by convolving the CLEAN component with the default beam shape of the GMVA.
}
\label{Fig:simulation1}
\end{figure}
\begin{figure}[H]
\begin{center} 
\epsscale{0.6}
\plotone{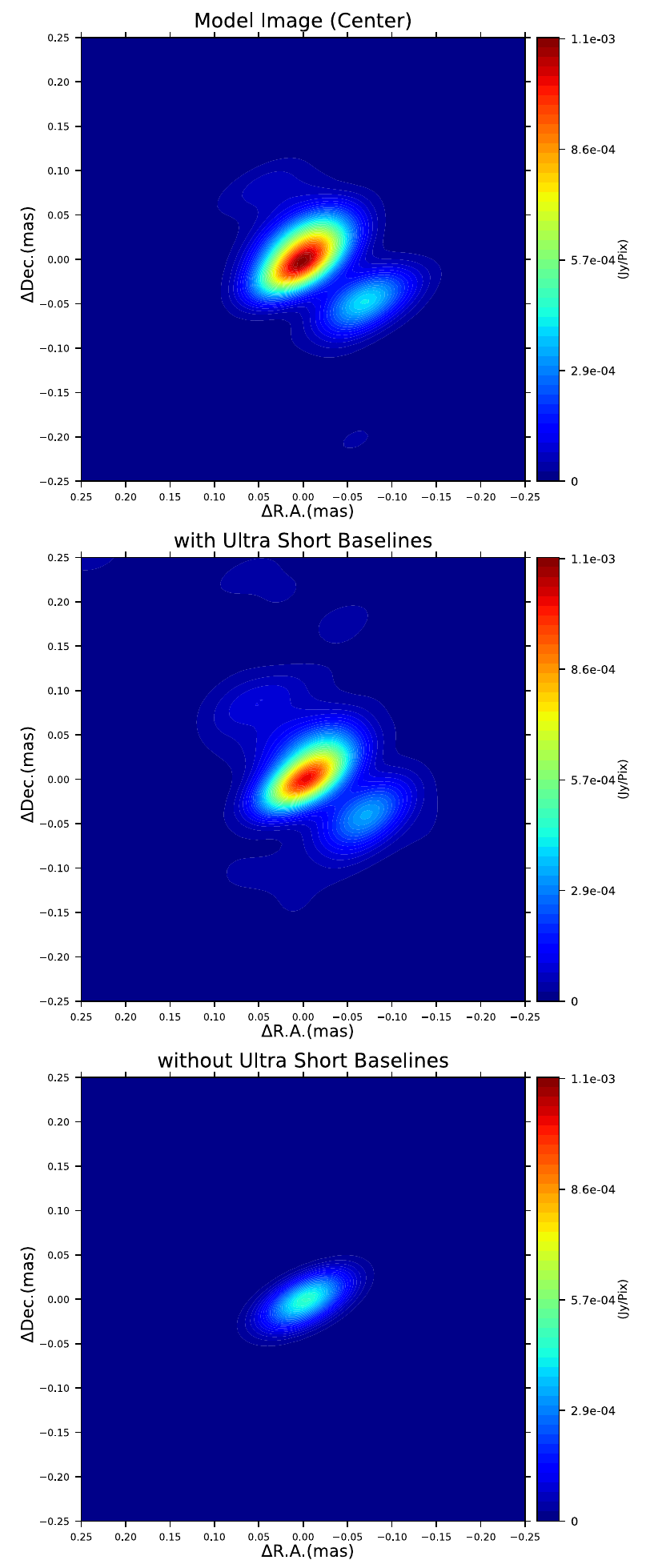}
\end{center} 
\vspace{-4mm}
\caption{Simulation for the performance of the data from the ultrashort base lines (II).Shown is the central core area. 
The top panel shows the model image, and the middle panel shows the result when all data including the ultrashort baseline data are used.
The bottom panel shows the result when the ultrashort baseline data are omitted and only the remaining long baseline data are used. The brightness distributions are generated by convolving the CLEAN component with the default beam shape of the GMVA.
}
\label{Fig:simulation2}
\end{figure}

\end{document}